# Real-Time Coupled Electron-Nuclear Dynamics of Chemical Bond Formation: Hydrogen Scattering from a Semiconductor Surface

Jialong Shi,[1] Lingjun Zhu,[1] Florian Nitz,[2] Oliver Bünermann,[2,3,4] Alec M. Wodtke,[2,3,4] Hua Guo,[5] and Bin Jiang[1,6*]

[1]*State Key Laboratory of Precision and Intelligent Chemistry, Department of Chemical Physics, University of Science and Technology of China, Hefei, Anhui 230026, China*

[2]*Institute of Physical Chemistry, Georg-August University, Göttingen 37077, Germany*

[3]*Department of Dynamics at Surfaces, Max-Planck-Institute for Multidisciplinary Sciences, Göttingen 37077, Germany*

[4]*International Center for Advanced Studies of Energy Conversion, Georg-August University, Göttingen 37077, Germany*

[5]*Department of Chemistry and Chemical Biology, Center for Computational Chemistry, University of New Mexico, Albuquerque, New Mexico 87131, USA*

[6]*Hefei National Laboratory, University of Science and Technology of China, Hefei 230088, China*

*: Corresponding author: bjiangch@ustc.edu.cn

**Abstract**

A first-principles coupled electron-nuclear dynamics simulation based on real-time, time-dependent density functional theory and Ehrenfest dynamics quantitatively reproduces bimodal translational energy loss and angular distributions observed in experiment for hydrogen atom scattering from Ge(111)-$c(2\times8)$. The theory elucidates a site-selective mechanism of electronically nonadiabatic energy transfer associated with the formation of different Ge-H bonds. When a hydrogen atom approaches a Ge rest-atom, it is strongly accelerated toward the potential minimum forming a transient Ge-H bond and then reflected by the repulsive wall. This transient bond formation triggers an ultrafast electron transfer event from the rest-atom to an adjacent Ge-adatom, involving several crossings between valence and conduction bands of the substrate. Electronic equilibration is impossible within such a short time (Born-Oppenheimer failure) allowing the H-atom kinetic energy to be converted to inter-band electronic excitation of the substrate. H-atom collisions at other Ge atoms also form a transient bond but exhibit no electronic excitation, resulting in distinctly less efficient energy loss in scattered H-atoms. The nuclear-to-electronic energy transfer observed in this system reflects the electronic dynamics of covalent bond formation at a semiconductor surface, a mechanism that is quite distinct from previously identified nonadiabatic energy transfer mechanisms at metal surfaces mediated by electronic friction or transient negative ions.

Chemical and photochemical reactions at surfaces govern a wide range of economically important processes, from heterogeneous catalysis to semiconductor device fabrication[1-3]. Yet, we are still far from understanding the simple act of forming a chemical bond at the gas-surface interface, which is an example of coupled electron-nuclear dynamics. When a bond forms, positively charged nuclei are brought into close contact, collectively attracting and stabilizing the bonding electrons between them. Excess energy must be dissipated to form a stable bond, either into nuclear or electronic motion, degrees of freedom that may be coupled to one another. Although the conventional density functional theory (DFT) enables a reasonable description of electronic ground states, current theoretical methods struggle to describe coupled electron-nuclear nonadiabatic dynamics in all but the simplest of systems.

One experimental approach to gaining relevant information about electron dynamics involved in bond formation is the scattering of atoms or molecules from surfaces, where measurements of translational and vibrational inelasticity[4, 5] reveal energy exchanges taking place between the projectile and the surface. In one study of H atom scattering at graphene surfaces, it was shown that formation of a transient C-H bond allowed large amounts of H-atom translational energy to be converted to phonon excitation[6]. Other studies have indicated that, in addition to energy dissipation into surface phonons, electronic excitation can be detected[7-10], suggesting the prospect that the electron-coupled nuclear dynamics of transient bond formation might be revealed in scattering experiments[5]. In fact, H-atom translational energy excites electrons in collisions with many metal surfaces[11-13], a nonadiabatic phenomenon without which the H atoms would not efficiently adsorb. These experiments have been explained by models involving molecular dynamics with electronic friction (MDEF)[12-15], implying that H-atom kinetic energy is lost in small chunks to many metal electrons.

Recently, experimental observation of H-atom scattering from a reconstructed semiconductor surface, Ge(111)-$c(2\times8)$, revealed clear evidence of nonadiabatic excitation of electrons from the valence band (VB) to the conduction band (CB)[16]. At low H-atom incidence energy ($E_i$), energy loss was small and the energy loss distribution of the scattered H atoms exhibited a single peak. However, for values of $E_i$ larger than the surface bandgap (~0.49 eV)[17], a bimodal energy-loss distribution was observed and the high-energy loss channel exhibited a negligible isotopic effect[18], providing experimental evidence for a site-specific non-adiabatic mechanism. While adiabatic molecular dynamics (MD) reproduced the low-energy loss peak, neither it nor MDEF could account for the high-energy loss component. These experimental observations reflect a collisional process that excites a single electron from the VB to the CB, for which new theoretical approaches are required.

A recent DFT-based theoretical study[19] provided clues to a possible explanation. In that work, strong mixings of Kohn-Sham (KS) orbitals belonging to VB and CB were found when the H atom was placed at certain energy minima corresponding to bond formation at specific surface sites. This suggested that a plausible hypothesis of the electronically nonadiabatic dynamics might involve the formation of a Ge-H bond. Verifying this hypothesis, however, requires a challenging first-principles simulation involving coupled electron-nuclear dynamics, one that is capable of describing the experimentally observed energy loss spectrum and angular distributions.

In this work, we have adapted a real-time time-dependent DFT (rt-TDDFT) method with Ehrenfest dynamics for the nuclear motion[20-22] allowing us to simulate the coupled electron-nuclear dynamics of this process on-the-fly. Our results reproduce experimental observations of H-atom kinetic energy loss and angular scattering distributions remarkably well. We find that the bimodal H-atom energy loss distribution observed in experiment arises from site-specific collisions at different Ge surface atoms present on the reconstructed surface. Specifically, the high energy loss feature arises from transient formation of a Ge-H bond at the Ge rest atom, which requires an ultrafast inter-site electron transfer. Contrasting this, transient bond formation at the Ge adatom and at saturated Ge atoms, exhibits no inter-site electron transfer, produces no electronic excitation and leads to only little energy loss. These results highlight the ability of rt-TDDFT to describe the starkly different ultrafast coupled electron-nuclear dynamics involved in formation of different types of covalent bonds occurring during H-atom collisions at these surface sites.

Within the framework of Ehrenfest dynamics governed by rt-TDDFT, nuclei evolve classically as governed by the average force generated by the time-dependent electronic density, while electrons evolve quantum mechanically as dictated by the time-dependent Kohn–Sham (TDKS) equations[20],

$$i\frac{\partial}{\partial t}\psi_i(\mathbf{r},t) = \widehat{H}[\rho(\mathbf{r},t)]\psi_i(\mathbf{r},t). \tag{1}$$

Here, $\psi_i(\mathbf{r},t)$ is the $i^{th}$ TDKS orbital and $\widehat{H}[\rho(\mathbf{r},t)]$ is KS Hamiltonian, which depends on the time-dependent electronic density $\rho(\mathbf{r},t)$. The mean-field force for the nucleus $F_I(\mathbf{R})$ arising from the weighted average of KS orbitals,

$$F_I(\mathbf{R}) = -\sum_{i=1}^{N_{occ}} |C_i(t)|^2 \nabla_I E_i - \sum_{\substack{i \neq j \\ }}^{N_{occ}} C_i^*(t)C_j(t)\langle\psi_i(\boldsymbol{R}|\boldsymbol{r},t)|\nabla_I\widehat{H}[\rho(\boldsymbol{r},t)]|\psi_j(\boldsymbol{R}|\boldsymbol{r},t)\rangle. \tag{2}$$

where $C_i(t)$ and $C_j(t)$ are coefficients of the $i^{th}$ and $j^{th}$ KS orbitals, respectively, $E_i$ is the $i^{th}$ eigenenergy. The right-hand side of Eq. (2) includes both the average adiabatic force and non-adiabatic force effectively involving couplings among occupied KS orbitals.

Although rt-TDDFT has been widely applied for photo-induced carrier dynamics[23, 24], its application to collisional dynamics has seldom been attempted[25, 26], especially to such an extended system. Specifically, we performed spin-polarized rt-TDDFT simulations with a hybrid density functional HSE06[27] and norm-conserving pseudopotentials[28], as efficiently implemented in PWmat[24, 29] that allows us to use a time step of 0.1 fs for the coupled electron-nuclear dynamics. More computational details are given in the supplementary information (SI).

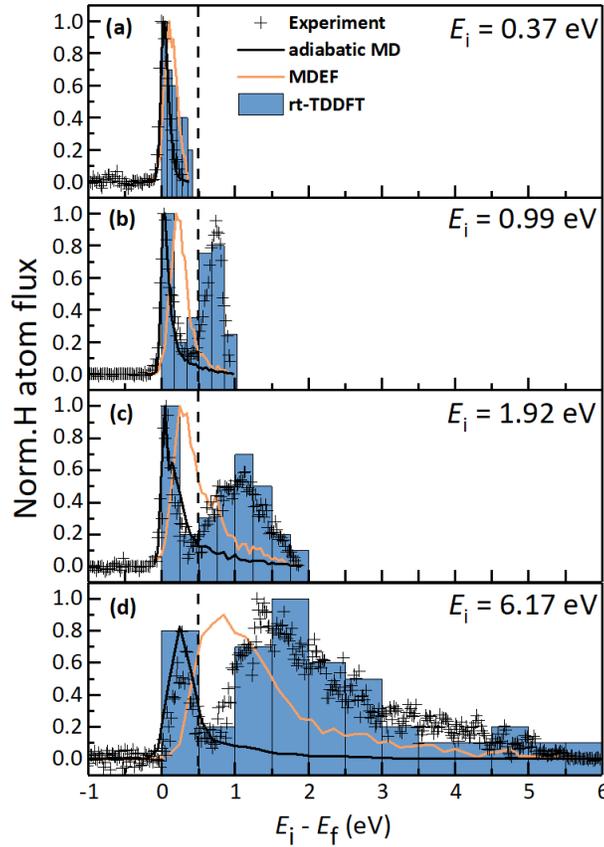

**FIG. 1. Comparison of translational energy-loss distributions for H atoms scattered from Ge(111)-c(2×8).** Experimental[16] (+), adiabatic MD[16] (black lines), MDEF[16] (orange lines) and rt-TDDFT (blue columns) results are compared at incidence energies of (a) 0.37 eV, (b) 0.99 eV, (c) 1.92 eV and (d) 6.17 eV. In rt-TDDFT simulations, the H atom was initialized at 4 Å above the surface with an incidence angle of 45° along the [$\bar{1}$10] surface direction. Due to the limited number of trajectories, all scattering events are counted in rt-TDDFT, while experimental data are derived from in-plane scattering flux only, for both of which the highest intensity is scaled to unity.

To make a quantitative comparison between theory and experiment, we have computed at least one hundred rt-TDDFT trajectories for each of four experimental conditions reported in Ref. 16, performing statistically meaningful sampling of initial conditions as detailed in SI and Fig. S1. It is worth highlighting that since the energy difference between the maximum of the VB (VBM) and the minimum of the CB (CBM) predicted by HSE06 (0.489 eV) agrees well with the measured surface bandgap of $0.49 \pm 0.03$ eV[17], no adjustment was required or made in our simulations.

As shown in Fig. 1a-d, only the calculated translational energy loss distributions obtained with rt-TDDFT agree with experimentally derived distributions; in fact, these agree remarkably well at four incidence energies from 0.37 to 6.17 eV. For incidence energies below the bandgap ($E_i$=0.37 eV), a single sharp energy loss peak is obtained; while for incidence energies above the bandgap, rt-TDDFT simulations reproduce the experimentally observed bimodal distributions, which exhibit a clean separation at the energy of the bandgap. The rt-TDDFT simulations also capture key characteristics of the high-energy-loss component, including its broadening and the shift of its peak position with increasing $E_i$.

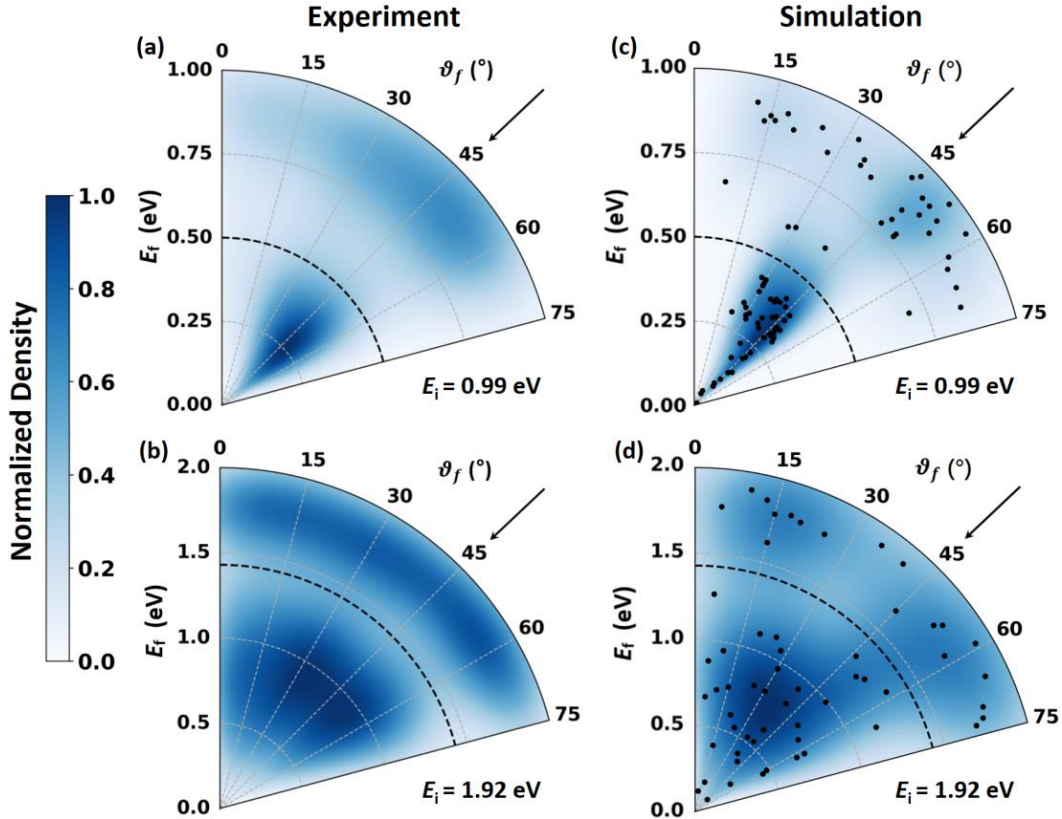

**FIG. 2. Angular distributions of H atoms scattered from Ge(111)-c(2×8).** Comparison of experimental (a, b) and rt-TDDFT simulation (c, d) results from an incidence angle of 45° along the [$\bar{1}$10] surface direction, at $E_i$=0.99 eV (upper panels) and $E_i$=1.92 eV (lower panels), both being kernel density distributions fitted to discrete scattering data; the raw theoretical data is shown as dots. Again, all scattering events are counted in rt-TDDFT, while experimental data are derived from in-plane scattering flux only.

The H atom scattering angular distributions at $E_i$=0.99 eV and $E_i$=1.92 eV obtained from rt-TDDFT also compare well with the experiment—see Fig. 2. At $E_i$=0.99 eV, the high energy loss channel exhibits a narrower angular distribution, compared to the low energy loss channel. At $E_i$=1.92 eV, both channels exhibit broader angular distributions.

The success of these first-principles simulations in reproducing experimental data is strong evidence that the rt-TDDFT with Ehrenfest dynamics accurately describes the coupled electron-nuclear dynamics of H scattering from a Ge(111)-$c(2\times8)$ surface This justifies using the rt-TDDFT framework to analyze the underlying scattering mechanism, a topic to which we now turn.

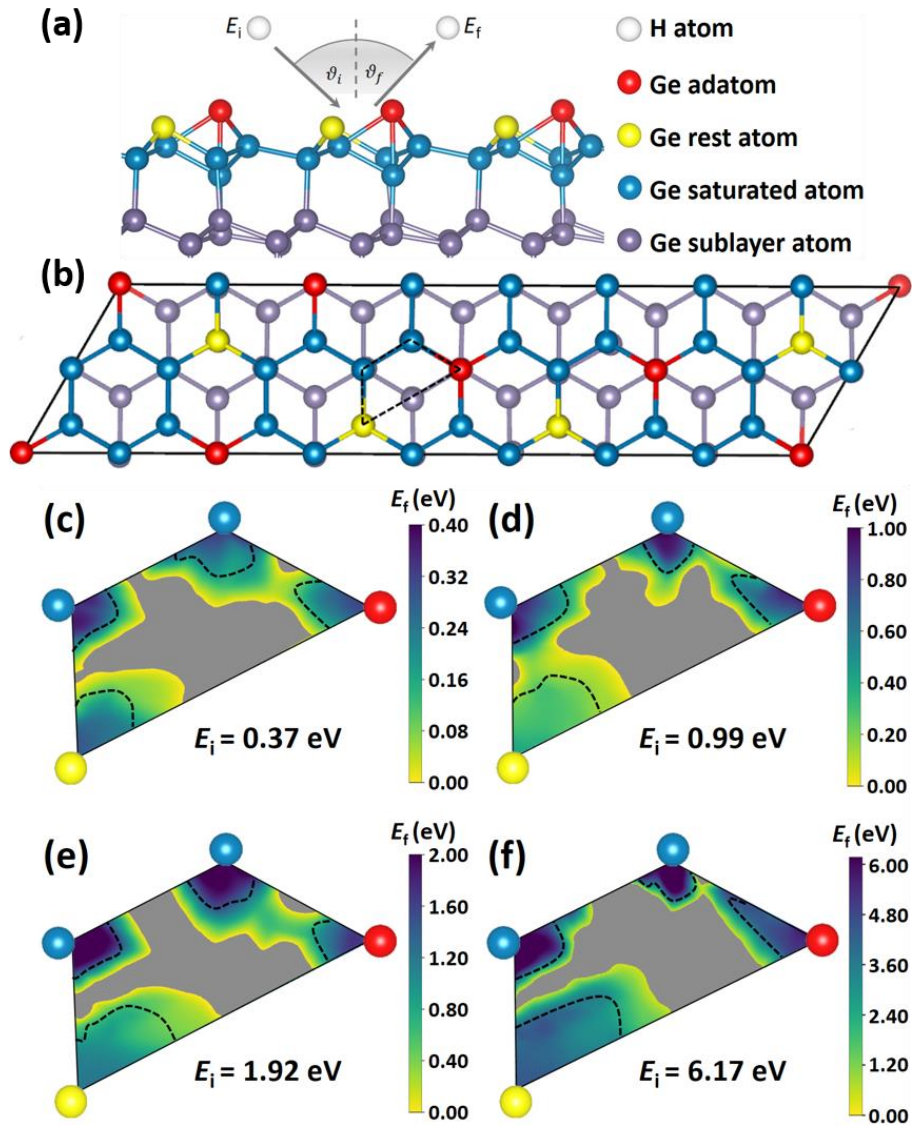

**FIG. 3. Ge(111) $c(2\times8)$ surface structure and dependence of H-atom scattering energy on initial impact point.** (a)

Schematic illustration of the scattering process with an atomic model of the reconstructed surface. (b) Top view of the surface. The smallest area that includes all surface sites is marked by dashed frame, from which initial impact positions were uniformly sampled. (c-f) Kernel density distributions of the scattered H-atom's final kinetic energy at four incidence energies (c) 0.37 eV, (d) 0.99 eV, (e) 1.92 eV and (f) 6.17 eV, where the colors reflect the energy scale. Dashed lines roughly encircle regions of direct scattering from specific Ge atoms. Gray areas indicate subsurface penetration and trapping.

The most significant finding is the strong site-specificity of the energy transfer process. To understand this, we must first consider the structure of the reconstructed Ge(111) surface. A hypothetical Ge surface structure formed by cleavage along the (111) direction is highly unstable, as it would exhibit many dangling bonds. The surface is stabilized by a reconstruction that can be illustrated in two steps. First, dangling bonds are eliminated in a structural reorganization to form a $c(2\times8)$ structure shown in Fig. 3a. This surface consists of 0.25 monolayer adatoms (shown in red) each possessing a single dangling bond and bound to three saturated atoms of the first layer (shown in blue). Additional dangling bonds are located at rest atoms in the first layer (shown in yellow). The surface is further stabilized by the transfer of one dangling bond electron from adatoms to rest atoms, which are present in equal quantities[30].

To illustrate the site-specific nature of the electronically nonadiabatic energy transfer process, Fig. 3 shows rt-TDDFT predictions of the H-atom scattering energy with respect to the initial impact point within a smallest unit that includes all surface sites. At $E_i$=0.37 eV where no electronic excitation is possible, H atoms that collide close to any Ge atom exhibit similar scattering energies. Interestingly, at higher incidence energies, H atoms that collide near the rest atom emerge with markedly reduced energy. Further analysis shows that large translational energy loss with associated electronic inter-band electron (VB-CB) excitation occurs via two types of events at all incidence energies: (1) direct scattering from the rest atom, and (2) collisions at a rest atom after an initial impact at another atom. These multiple-bounce collisions become increasingly important at higher incidence energy. At $E_i$=6.17 eV, they are responsible for the long tail of the energy loss distribution stretching beyond ~3 eV. In stark contrast to this behavior, direct scattering from isolated adatoms or saturated atoms results in much less energy loss, reflecting electronically adiabatic behavior. These adiabatic processes contribute mainly to the low-energy-loss peak in the energy loss distributions.

Note also that about half of incident H atoms collide between Ge atoms and penetrate the surface. These H atoms undergo multiple collisions with sub-surface Ge atoms and do not escape within the simulation timeframe (up to 300 fs). The implied trapping probability is ~56% at $E_i$=0.37 eV and decreases very slightly with increasing incidence energy. Since these trapped trajectories are undetectable in scattering experiments, they are excluded from the energy loss analysis.

We can also use the rt-TDDFT to distinguish energy dissipation to surface phonons from that into EHPs. Analyzing trajectories for $E_i$=1.92 eV as an example, we find that H-atom collisions at all surface sites are capable of exciting surface phonons, leading to an increase of kinetic energy of the substrate Ge atoms after scattering (Fig. S2). The mean adiabatic energy loss to phonons is 0.24 eV. Only H-atom collisions involving the rest atom—direct or indirect—excite electrons to the CB through coupled electron-nuclear dynamics. These nonadiabatic events lead to a mean electronic excitation of 0.86 eV.

The site-specific behavior supports the hypothesis presented above—electronically nonadiabatic dynamics involve formation of a Ge-H bond. Specifically, we see that promotion of electrons above the band gap requires the H atom to collide with a Ge rest atom, where a Ge-H bond can form. On the other hand, Ge-H bond formation is also possible at the adatom or at saturated atoms, but does not lead to electronic excitation. We next explain this site-specific behavior by showing that chemical bond formation at the rest atom is quite different than at the adatom.

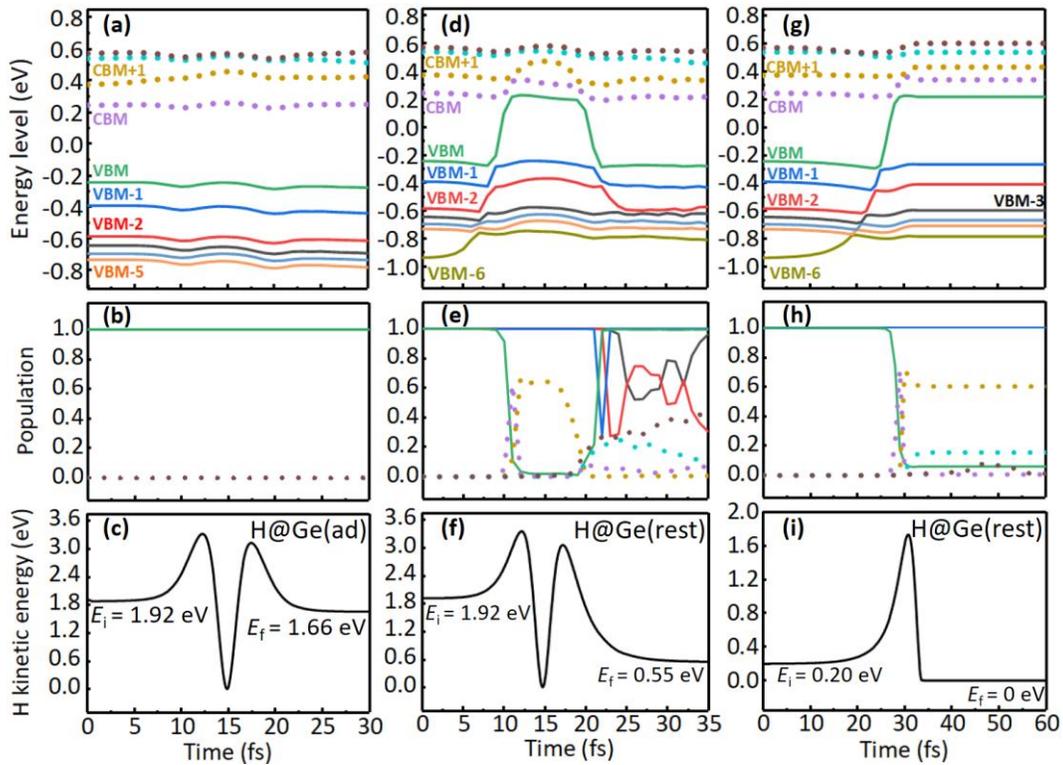

**FIG. 4. Representative trajectories of H scattering from Ge(111)-$c$(2×8) using rt-TDDFT with Ehrenfest dynamics.** Temporal evolution of KS orbital energies (a, d, g), electronic populations of these orbitals (b, e, h), and H-atom kinetic energy (c, f, i) of exemplary trajectories of H-atom scattering from a Ge adatom (a, b, c) and a Ge rest-atom (d, e, f) at

$E_i$=1.92 eV, and from a Ge rest-atom at $E_i$=0.20 eV (g, h, i). Solid and dotted lines denote VB and CB energy levels, which are referenced to the Fermi level of the initial configuration and labeled by their energy closeness to the VBM and CBM, respectively. The surface was initiated at 0 K to avoid thermal mixing of KS orbitals before impact, thereby simplifying the dynamical analysis.

To see this, Fig. 4(a-i) shows the coupled electron-nuclear dynamics of selected trajectories obtained from rt-TDDFT with Ehrenfest dynamics. For simplicity, we direct an impinging H atom along the surface normal toward either an adatom or a rest atom. H collisions at an adatom site at $E_i$=1.92 eV shown in panels 4(a-c) experience rapid acceleration to the surface indicating formation of a transient Ge-H bond; however, they do not excite electrons. The VBM populations (Fig. 4b) do not change during the collision, and the collision induces only small changes to the energies of the KS energy levels (Fig. 4a). This shows clearly that H-atom collisions at adatoms reflect ground electronic state dynamics exciting only surface phonons ($\langle \Delta E \rangle \approx 0.26$ eV) to energies less than that of the bandgap. Similar behavior is seen at other values of $E_i$ indicating that this adiabatic mechanism is general at the adatom (Fig. S3).

H-atom collisions at Ge rest-atoms (*e.g.*, $E_i$=1.92 eV) behave quite differently as the band structure is split into two spin-polarized manifolds, as discussed previously[19], one of which undergoes inter-band transitions (Fig. 4d). Beginning at $t \approx 10$ fs, the trajectory encounters an avoided crossing between VBM and CBM followed by another avoided crossing between CBM and CBM+1 and transient population is observed in CBM, which is then transferred to CBM+1—see Fig. 4e. Near the H-atom's turning point at $t \approx 15$ fs, the VBM has become completely depopulated forming an EHP (CBM+1, VBM). With subsequent acceleration of the H atom leaving the surface, additional electronic transitions are seen involving rapid population transfer between sub-bands, each correlated with its own avoided crossing. After the H atom has departed (>25 fs), the VBM (CBM+1) has become fully repopulated (depopulated), and two EHPs (CBM+3/CBM+2, VBM-2/VBM-3) with similar energies have been created. Overall, the excited electrons are unable to keep up with the nuclear motion within a short time and the coupled electron-nuclear dynamics result in a ~1.37 eV loss of H-atom translational energy (Fig. 4f). Similar behavior is found for trajectories initialized at $E_i$=0.99 eV (Figs. S3 and S4). Note that the other spin manifold shows no curve crossings and behave adiabatically—see Fig. S4.

It is interesting to compare the results just described to those of a trajectory at $E_i$=0.20 eV incident at a rest atom, depicted in Fig. 4(g-i). Here, the incident H atom is accelerated by bond formation to a maximum kinetic energy of ~1.8 eV, which is more than sufficient to promote inter-band electronic excitation just as in the higher incidence energy trajectory. However, unlike the high-energy scattering events, there is no recoil. Here, the H atom remains bound to the Ge rest-atom dissipating its entire kinetic energy in an astonishingly short time (~3 fs), confirming that electronic excitation is involved in stabilizing the newly formed bond between the H and the Ge rest-atom. Specifically, one sees that CBM+1 and, to a lesser degree CBM+2, are now populated—note that these are the same states that are transiently populated in high energy trajectories. This comparison makes clear that the electronically nonadiabatic energy transfer at high $E_i$ proceeds through transient population of the same sub-bands involved in formation of the chemical bond between the H atom and the Ge rest atom.

These observations stand in contrast to the electronically adiabatic behavior seen for an H-atom collision and transient H-Ge bond formation at an adatom. The explanation for this difference in behavior resides in the qualitative difference in the bond formation motif possible at the rest and the adatom. As mentioned above, the $c(2\times8)$ reconstruction involves an electron transfer from the adatom to the rest atom. Hence, H atom's attack at the adatom involves overlap of the singly occupied H atom 1s orbital with the unoccupied Ge $4p_z$ orbital. The resulting bonding orbital remains occupied by a single electron and leads to electronically adiabatic scattering. By contrast, H atom attack at a Ge rest atom involves three electrons, one from the H 1s orbital and two from Ge $4p_z$ orbital. In order to form a 2-electron Ge-H bond, the third electron must be promoted to an unoccupied orbital above the Fermi energy—this can be seen in Fig. 4(d, e) and Fig. 4(g-h).

To illustrate this more clearly, Fig. 5(a-f) shows the time-dependence of the projected density of *occupied* states (PDOS) onto the H-1s orbital, $4p_z$ orbitals of the Ge rest atom (Ge(rest)-$4p_z$) and adatom (Ge(ad)-$4p_z$) along a representative trajectory impacting a rest atom at $E_i$ =1.92 eV. Also displayed are the electron density distributions of most relevant orbitals ($|\psi_i(\mathbf{r},t)|^2$), corresponding to the resonance energy of the isolated H and Ge orbitals when H is far from the surface and of the bonding and antibonding orbitals when Ge-H bond temporarily forms. At the beginning of the trajectory, the non-interacting H-1s and Ge(rest)-$4p_z$ orbitals appear respectively near –4 eV and –1 eV, while the Ge(ad)-$4p_z$ orbital is largely unoccupied with a negligible PDOS. By ~8 fs, H-Ge bonding interactions clearly arise as H-1s and Ge(rest)-$4p_z$ resonances start to overlap, forming bonding and anti-boding orbitals (both occupied). As the H-Ge interaction strengthens, the splitting of the resonances increases and the associated electron density largely shifts from the rest atom to an adjacent adatom, stabilizing the two-electron H-Ge(rest) bond by moving the extra electron to states localized on the Ge adatom (11 fs and 15 fs). This is reminiscent of Hoffmann's Type-3 bonding[31] and results in electronic excitation from VB to CB. As the H atom scatters back to vacuum, the electron density shifts back from the adatom to rest atom (22 fs), recovering the isolated H-1s and Ge(rest)-4p$_z$ orbitals and leaving excited surface EHP(s) (28 fs). This reversible charge transfer—an inter-band transition—occurs within ~20 fs.

Fig. 5(g-l) shows the same PDOS plots for an H atom collision at an adatom. Here the inter-site electron transfer does not take place. Instead, one can see that H-1s hybridizes with the Ge(ad)-4p$_z$ orbital, forming a transient one-electron H-Ge(ad) bond involving the resonance at about −4.5 eV. Note that during this trajectory at no time is there orbital population above the Fermi energy at $t=0$. This reflects the detailed electronic characteristics of an adiabatic trajectory taking place on the ground electronic state.

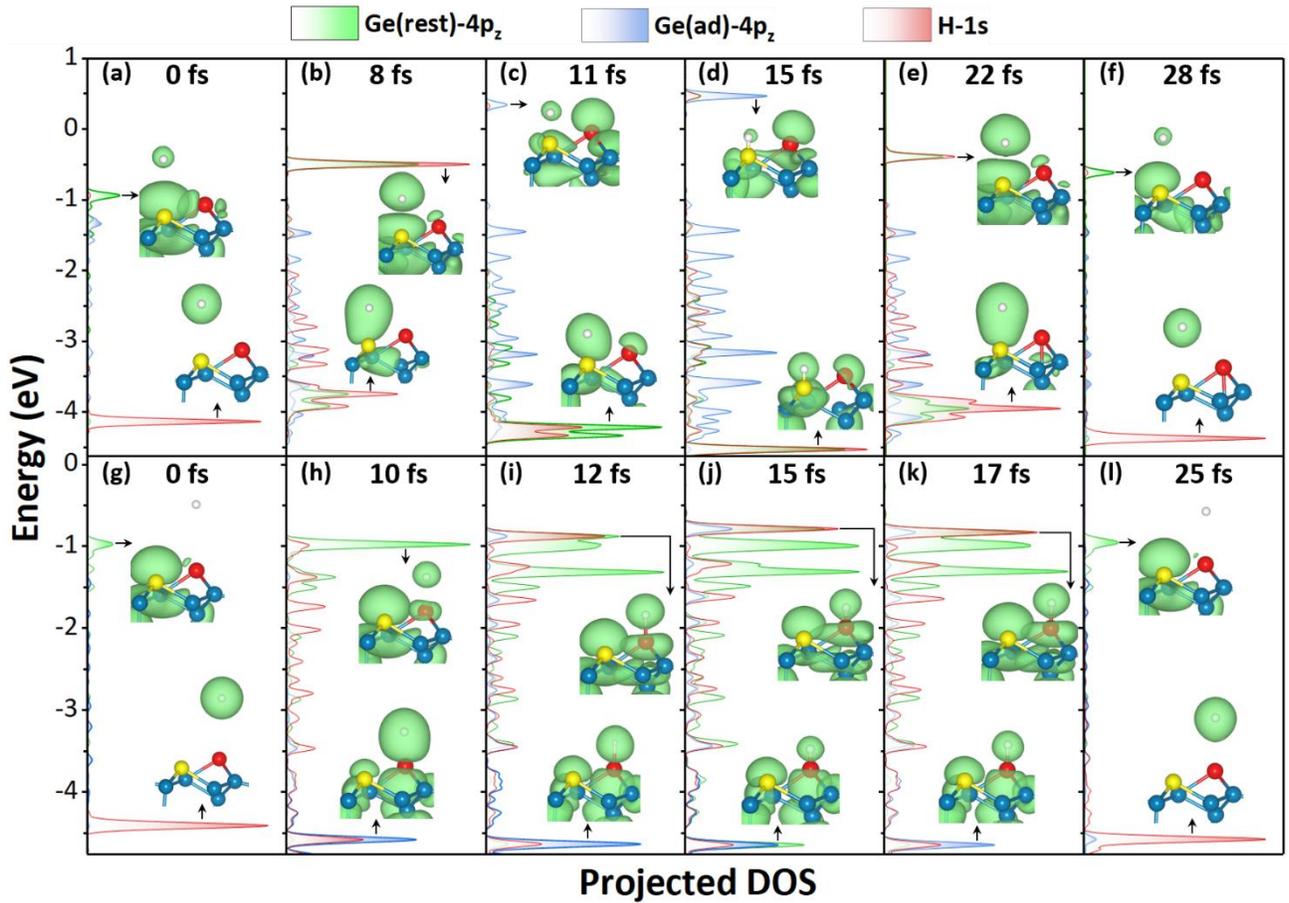

**FIG. 5. Temporal evolution of PDOSs and orbital-specific electron densities.** Snapshots of the PDOSs of H-1s and Ge-4p$_z$ orbitals during the representative scattering trajectories of Fig. 4 (c,f) are shown for an H-atom collision at a Ge rest-atom (a-f) and at a Ge adatom (g-l) at $E_i$=1.92 eV. Also shown as insets are the time-dependent orbital densities ($|\psi_i(r,t)|^2$). At time=0 & 25 fs, the separated H-1s and Ge(rest)-4p$_z$ orbitals are seen. The dominant components of the bonding and antibonding orbitals can be seen at intermediate times. In each row of panels, the energy zero is aligned to the Fermi level of the initial configuration (at $t=0$ fs), where H is far from an equilibrium Ge surface. See text for details.

The near quantitative agreement between rt-TDDFT with Ehrenfest dynamics and experiment has made a clear elucidation of the coupled electron nuclear dynamics of the covalent bond formation processes possible in this system. We wish to emphasize several key factors that we believe led to the success of this approach. First, it appears that the rt-TDDFT framework correctly captures the avoided crossings and nonadiabatic couplings between KS orbitals along the scattering trajectory as the surface atomic structure barely changes, despite the surface band structure being dramatically altered during collision. Second, the hybrid functional HSE06 provides reliable energetics—it is not likely that lower level functionals would have performed so well in this study. Third, the ultrafast nature of this process and the dense manifolds of electronic states mitigate some limitations of Ehrenfest dynamics, such as its known violation of detailed balance in long-time simulation. It is noteworthy that Ehrenfest dynamics coupled with a Haldane-Anderson model failed to reproduce experimental observations[32]. This contrast underscores that an accurate, first-principles description of the electronic structure—as achieved here with rt-TDDFT and a hybrid functional—is essential for quantitatively modeling of nonadiabatic dynamics at semiconductor surfaces. Admittedly, the theory-experiment agreement is not perfect, likely due to the mean-field approximation or other simplifications necessitated by computational constraints. For instance, the chosen area for initial sampling underestimates contributions from saturated atoms, which likely affects the electronically adiabatic scattering. This may account for minor discrepancies in the energy loss distribution at $E_i =$ 0.37 eV and angular distributions of the adiabatic channel as seen in Figs. 1 and 2. We have also used all scattering trajectories to construct the energy loss distribution due to the heavy computational costs, whereas the experiment only observes scattering within a single scattering plane. Addressing these issues represents a valuable direction for future work.

In conclusion, we have presented here results from rt-TDDFT simulations combined with Ehrenfest dynamics with adequate initial condition sampling to understand experimental observations for H atoms scattered from a Ge(111)-$c$(2×8). The results obtained from this first-principles model are in excellent agreement with experiment, allowing us to gain insights into the electronic dynamics occurring when an H atom collides with a semiconductor surface. The analysis of time-resolved orbital density distributions along nuclear trajectories reveals an ultrafast, collision-induced inter-site electron transfer associated with formation of a transient chemical bond at a Ge-rest atom, that triggers a cascade of electronic band crossings allowing H atom kinetic energy to promote an electron to the CB. These are the same electronic excitations involved in stabilizing a newly formed H-Ge bond. Collisions at the Ge adatom also form a transient bond, but exhibit no inter-site electron transfer or electronic excitation. The observations and insights made in this work reveal a new nonadiabatic energy transfer mechanism intimately related to the nature of covalent bond formation on semiconductor surfaces. It differs fundamentally from the EHP excitation commonly invoked for nonadiabatic processes at metal surfaces, which have been explained by electronic friction[12-15] or transient negative ion formation[33-42].

**Acknowledgements**: This work is supported by the Innovation Program for Quantum Science and Technology (2021ZD0303301 to B.J.) and the National Natural Science Foundation of China (22325304 and 22221003 to B.J.). H.G. acknowledges support from the US National Science Foundation (Grant. No. CHE-2306975) and the Alexander von Humboldt Foundation for a Humboldt Research Award. F.N. acknowledges support from the BENCh graduate school, funded by the Deutsche Forschungsgemeinschaft (DFG, German Research Foundation) 389479699/GRK2455. OB and AW acknowledge support from the Deutsche Forschungsgemeinschaft (DFG, German Research Foundation) under grant number 217133147/SFB 1073, project A04.